\documentclass{ws-mpla}
\usepackage[normalem]{ulem}  % \sout{old text} for strikeout
\usepackage[dvips]{color} % For blue in-text comments and additions
\renewcommand\sout{\bgroup \color{red} \ULdepth=-.5ex \ULset}

\begin{document}

\markboth{S. Yasui, S. H. Lee, K. Ohnishi,  I.-K. Yoo and C. M. Ko}
{STUDYING DIQUARK STRUCTURE OF HEAVY BARYONS IN RELATIVISTIC
HEAVY ION COLLISIONS}

%%%%%%%%%%%%%%%%%%%%% Publisher's Area please ignore %%%%%%%%%%%%%%
\catchline{}{}{}{}{}
%%%%%%%%%%%%%%%%%%%%%%%%%%%%%%%%%%%%%%%%%%%%%%%%%%%%%%%%%%%%%%%%%%%

\title{STUDYING DIQUARK STRUCTURE OF HEAVY BARYONS IN RELATIVISTIC
HEAVY ION COLLISIONS}

\author{\footnotesize S. Yasui}

\address{Institute of Physics
and Applied Physics, Yonsei University, Seoul 120-749, Republic of Korea \\
yasui@phya.yonsei.ac.kr\\
Department of Physics, National Taiwan University,
  Taipei, 106, Taiwan\\
yasui@phys.ntu.edu.tw}

\author{\footnotesize S. H. Lee, K. Ohnishi}

\address{Institute of Physics and Applied Physics, Yonsei University,
Seoul 120-749, Republic of Korea \\
suhoung@phya.yonsei.ac.kr\\
kohnishi@phya.yonsei.ac.kr}

\author{\footnotesize I.-K. Yoo}

\address{Pusan National University, Pusan 609-735, Republic of Korea\\
yoo@pusan.ac.kr}

\author{\footnotesize C. M. Ko}

\address{Cyclotron Institute and
Physics Department, Texas A\&M University, College Station, TX
77843, U.S.A.\\
ko@comp.tamu.edu}

\maketitle

\pub{Received (Day Month Year)}{Revised (Day Month Year)}

\begin{abstract}
We propose the enhancement of $\Lambda_c$ yield in heavy ion
collisions at RHIC and LHC as a novel signal for the existence of
diquarks in the strongly coupled quark-gluon plasma produced in
these collisions as well as in the $\Lambda_c$. Assuming that stable
bound diquarks can exist in the quark-gluon plasma, we argue that
the yield of $\Lambda_c$ would be increased by two-body
collisions between $ud$ diquarks and $c$ quarks,
in addition to normal three-body collisions among
$u$, $d$ and $c$ quarks. A quantitative study of this effect based
on the coalescence model shows that including the contribution of
diquarks to $\Lambda_c$ production indeed leads to a substantial
enhancement of the $\Lambda_c/D$ ratio in heavy ion collisions.

\end{abstract}

\ccode{PACS Nos.: 12.38.Mh, 14.20.Lq, 12.38.Qk}

%\section{Introduction}

Recent experiments on heavy ion collisions at the Relativistic Heavy
Ion Collider (RHIC) have shown that the properties of 
produced quark-gluon plasma (QGP) are close to those of a perfect
fluid \cite{Teaney:2000cw}, consistent with the strongly
coupled nature of QGP suggested in
Ref. \refcite{Policastro:2001yc}. The strong correlations in the QGP
may give rise to a rich variety of states
\cite{Datta:2002ck,Asakawa:2003re}.
% For example, the lattice QCD simulation has indicated that the $c\bar{c}$ bound state
%could survive in the QGP up to several times the critical
%temperature \cite{Datta:2002ck,Asakawa:2003re}.
This new picture of the QGP as a strongly coupled system is now
called the strongly coupled QGP (sQGP)
\cite{Asakawa:2003re,Hatsuda:1985eb,Shuryak:2003ty}. One of the
challenging problems in future relativistic heavy ion collisions is
to experimentally study such strong correlations. Recently,
 we have shown that possible existence of diquark correlations
in sQGP, particularly the $[ud]$ diquark with color $\bar{\bf
3}_{\rm c}$, isospin 0, and spin 0, would lead to a
significant enhanced production of $\Lambda_c$ in relativistic heavy
ion collisions compared to that of charmed mesons \cite{Lee:2007wr},
thus offering one possible experimental means to verify the
existence of strong correlations in sQGP.

The basic idea in our study is the following. If $[ud]$ diquarks can exist in the QGP
from the heavy ion collisions, the decoupling of the $[ud]$
diquark from the heavy quark in the heavy baryon
\cite{Jaffe:2004ph} would then allow the $[ud]$ structure to be
preserved during the hadronization process, thus facilitating the
production of $\Lambda_{c}$ ($\Lambda_{b}$) from the QGP. In other
words, in the presence of diquarks $[ud]$ the $\Lambda_{c}$
($\Lambda_{b}$) can be formed from the two-body collision between
$c$ ($b$) quark and the $[ud]$ diquark. Compared to the usual
three-body collision among $c$ ($b$), $u$ and $d$ quarks, the
two-body collision has a larger phase space than the
three-body collision. Therefore, collisions of $[ud]$ diquarks and
heavy $c$ ($b$) quarks in the QGP would enhance the yield of
$\Lambda_{c}$ ($\Lambda_{b}$) in heavy ion collisions.
%In the following, we first estimate the yield of $\Lambda_{c}$ in
%heavy ion collisions experiments at RHIC and LHC, and then
%generalize the study to the yield of $\Lambda_{b}$.

To estimate the yield of $\Lambda_{c}$, we have used the
quark coalescence model
\cite{Voloshin03,Hwa:2002tu,greco,Fries:2003vb}. For heavy ion
collisions, the coalescence model has been quite successful in
describing the pion and proton transverse momentum spectra at
intermediate momenta as well as at low momenta if resonances are
included \cite{greco1}. It also gives a natural account for the
observed constituent quark number scaling of hadron elliptic flows
\cite{kolb} and the large elliptic flow of charmed mesons
\cite{greco2}. In this model, the yield of hadrons from the QGP is
given by the overlap between the quark thermal distribution in the
QGP and the hadron wave function \cite{chen,Chen:2003tn}. Modeling
the formed QGP 
% in heavy ion collisions at RHIC and LHC
by a fire cylinder, the number of light quarks at hadronziation is
then determined from assuming that they are in thermal and chemical
equilibrium.
%For the $c$ quark, its
For $c$ quarks, their number is
estimated from initial hard nucleon-nucleon collisions
\cite{chen,Zhang:2007dm}, and they are also taken to be
thermal equilibrium as they seem to interact strongly in QGP as
well \cite{van Hees:2004gq,Zhang:2005ni,Molnar:2004ph}.

\begin{figure}[tb]
\centerline{\psfig{file=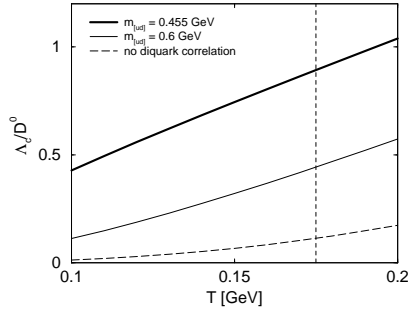,width=2.1in}}
\vspace*{8pt} \caption{The yield ratio $\Lambda_{c}/D^{0}$ as a
function of temperature. See the text.
% The bold and thin solid lines are from
%results for diquarks with masses $m_{\rm[ud]}=0.455$ GeV and 0.6
%GeV, respectively. The dashed line corresponds to the case without
%the diquark correlation. The vertical dashed line denotes $T_{\rm
%C}=0.175$ GeV.
\protect\label{fig1}}
\end{figure}

Results on the yield of $\Lambda_{c}$ is plotted in Fig. 1 as a
function of the temperature of QGP. The dashed line corresponds to
the case without diquarks, while solid lines are for the
cases with diquarks of masses $m_{[ud]}=450$ MeV (bold solid line)
or 600 MeV (thin solid line). The diquark mass is estimated by
assuming a constituent light quark mass of 300 MeV and taking into
account the color-spin interaction between the two quarks
\cite{Lee:2007tn}. The diquark mass of 450 MeV corresponds to the
maximum binding energy (150 MeV), while the diquark mass 600 MeV
corresponds to the minimum binding energy at threshold. Taking the
critical temperature to be $T_{\rm C}=175$ MeV, we obtain
$\Lambda_{c}/D^{0}=0.11$ for the case without diquarks and
$\Lambda_{c}/D^{0}=0.44$ and $0.89$ for the cases with diquarks of
masses 450 MeV and 600 MeV, respectively. Including diquarks in the
QGP thus enhances the $\Lambda_{c}/D^{0}$ ratio in relativistic
heavy ion collisions by a factor of 4 to 8 depending on how deeply
the diquark is bounded in QGP.

%\section{Discussions}

The predicted $\Lambda_{c}/D^{0}$ ratio is much larger
than that given by the statistical model \cite{PBM07}, which
is estimated to be
$(m_{\Lambda_c}/m_{D^0})^{3/2}\exp(-(m_{\Lambda_{c}}-m_{D^0})/T_{\rm
C}) \simeq 0.12$ \cite{Fries:2003vb}. It is also larger than the
value $\Lambda_{c}/D^{0}=0.159$ extracted from hadron
yields in p+p collisions by the SELEX collaboration
\cite{Kushnirenko:2000ed} and the value $\Lambda_{c}/D^{0}=0.14$
determined from measured hadron yields from
 $B$ meson decay \cite{Yao:2006px}. Since the $u$ and $d$ quarks in above three
cases are not correlated as a diquark, it is
therefore not surprising that the $\Lambda_{c}/D^{0}$ ratios
in these cases are all close to the value $\Lambda_{c}/D^{0}=0.11$
obtained from the coalescence model without diquarks in the QGP and the statistical
hadronization model of QGP for heavy ion collisions.

%The $\Lambda_{c}'s$ produced in QGP may change into $D$ mesons in
%the hadronic phase due to  collisions such as $\Lambda_{c} \pi
%\rightarrow N D ( D^{\ast})$. With the pion threshold momentum
%$p_{\rm th} \simeq 0.43$ GeV in the $\Lambda_{c} \pi \rightarrow N
%D$ process, which is larger than the typical energy scale $T_{\rm
%C}$, the life time due to the process is estimated as
%\begin{eqnarray}
%1/\tau = \Gamma_{\rm th} = 3 \int_{p_{\rm th}}^{\infty} \sigma n(p)
%\frac{{\rm d}^{3}p}{(2\pi)^{3}}
%\end{eqnarray}
%with 3 being the isospin factor, $\sigma$ the cross section, and
%$n(p)\simeq \exp(-\sqrt{p^{2}+m_{\pi}^{2}}/T)$. With $\sigma=5$ mb
%as a reasonable value suggested from the $J/\psi$ dissociation
%\cite{Liu:2001ce} and $T=T_{\rm C}$ for simplicity, we obtain $\tau
%\simeq 17.8$ fm, which is comparable with the lifetime of the
%hadronic phase $t_{\rm H}\simeq 10$ fm \cite{Chen:2003tn}, leading
%thus to a suppression factor $e^{-t_{\rm H}/\tau} \simeq 0.57$ for
%the $\Lambda_{c}$ yield. Since the temperature in the hadronic phase
%is lower than $T_{\rm C}$, the actual suppression factor will be
%closer to one. Therefore, the predicted $\Lambda_{c}$ enhancement in
%relativistic heavy ion collisions due to the presence of diquarks in
%the produced QGP is expected to survive final-state hadronic
%processes.

The study of charmed hadron production at RHIC and LHC will be
possible in the near future as the tracking system for observing
charmed hadrons, which has a spatial resolution of 12 $\mu$m with
the best precision, is being developed by the ALICE collaboration at
LHC \cite{alice1}. Also, detector upgrades
are in progress for both the STAR and
PHENIX collaborations at RHIC \cite{STAR1,PHENIX1}.

Although the abundance of $b$ quarks produced in heavy ion
collisions is much smaller than that of $c$ quarks, bottom hadrons
are also an interesting probe of the diquark correlation in
produced QGP. First, the decoupling between the diquark
and heavy quark holds even better as the heavy quark becomes
massive. Second, the lifetime of $\Lambda_b$ ($c\tau \sim 372$
$\mu$m) is longer than that of $\Lambda_{c}$ ($c\tau \sim 62$
$\mu$m), and this makes its detection easier. Using the bottom quark
production cross sections predicted from the pQCD for p+p collisions
at RHIC \cite{vogt1} and LHC \cite{vogt2}, we have estimated the
bottom quark numbers in corresponding heavy ion collisions. The
coalescence model then gives $\Lambda_{b}/B=0.098$ for the case
without diquarks, and $\Lambda_{b}/B=0.38$ and $0.82$ for the cases
with diquarks of masses 450 MeV and 600 MeV, respectively. The
enhancement of the $\Lambda_{b}/B$ ratio due to the presence of
diquarks is thus comparable to that for the $\Lambda_{c}/D$ ratio.

%\section{Summary}

%We have shown in this talk that.
In summary, the $[ud]$ diquark correlation in
the QGP produced in relativistic heavy ion collisions could be
investigated through the observation of $\Lambda_{c}$ and
$\Lambda_{b}$ enhancement in these collisions.
% Based on the picture
%that the $[ud]$ diquark is almost independent of the heavy quark in
%$\Lambda_{c}$ ($\Lambda_{b}$), the yield of $\Lambda_{c}$
%($\Lambda_{b}$) from the QGP is enhanced by the two-body collision
%between $c$ ($b$) quark and $[ud]$ diquark, in addition to the
%three-body collision among $c$ ($b$), $u$ and $d$ quarks.
Our proposal may open a possible new method
 to investigate the formation of QGP in relativistic heavy
ion collisions and to study the diquark correlation in the
QGP. It can in turn also be used as an
indirect evidence for the diquark structure in heavy baryons.
%Our discussions
%are thus related to many aspects of both hadron and quark physics.
We therefore conclude that studying $\Lambda_{c}$ and $\Lambda_{b}$
production at RHIC and LHC is an interesting and exciting subject.

%\section*{Acknowledgments}

The work was supported by the Korea Research Foundation
KRF-2006-C00011 (SHL, KO, SY and IKY), the Korea Science and
Engineering Foundation R01-2005-000-10334-0 (IKY), the US
National Science Foundation under Grant No.\ PHY-0457265 and the
Welch Foundation under Grant No.\ A-1358 (CMK), and the National Science Council of Taiwan under Grant No. NSC96-2112-M002-021 (SY).


\begin{thebibliography}{0}

\bibitem{Teaney:2000cw}
   D.~Teaney, J.~Lauret and E.~V.~Shuryak,
  %``Flow at the SPS and RHIC as a quark gluon plasma signature,''
  Phys.\ Rev.\ Lett.\  {\bf 86}, 4783 (2001);
%  [arXiv:nucl-th/0011058].
  %%CITATION = PRLTA,86,4783;%%
%\bibitem{Kolb:2000fh}
    P.~F.~Kolb, P.~Huovinen, U.~W.~Heinz and H.~Heiselberg,
  %``Elliptic flow at SPS and RHIC: From kinetic transport to  hydrodynamics,''
  Phys.\ Lett.\  B {\bf 500}, 232 (2001).
%  [arXiv:hep-ph/0012137].
  %%CITATION = PHLTA,B500,232;%%

\bibitem{Policastro:2001yc}
   G.~Policastro, D.~T.~Son and A.~O.~Starinets,
  %``The shear viscosity of strongly coupled N = 4 supersymmetric Yang-Mills
  %plasma,''
  Phys.\ Rev.\ Lett.\  {\bf 87}, 081601 (2001).
%  [arXiv:hep-th/0104066].
  %%CITATION = PRLTA,87,081601;%%

\bibitem{Datta:2002ck}
  S.~Datta, F.~Karsch, P.~Petreczky and I.~Wetzorke,
  %``A study of charmonium systems across the deconfinement transition,''
  Nucl.\ Phys.\ Proc.\ Suppl.\  {\bf 119}, 487 (2003).
%  [arXiv:hep-lat/0208012].
  %%CITATION = NUPHZ,119,487;%%

%\cite{Asakawa:2003re}
\bibitem{Asakawa:2003re}
  M.~Asakawa and T.~Hatsuda,
  %``J/psi and eta/c in the deconfined plasma from lattice QCD,''
  Phys.\ Rev.\ Lett.\  {\bf 92}, 012001 (2004).
%  [arXiv:hep-lat/0308034].
  %%CITATION = PRLTA,92,012001;%%

%\cite{Hatsuda:1985eb}
\bibitem{Hatsuda:1985eb}
  T.~Hatsuda and T.~Kunihiro,
  %``Fluctuation Effects In Hot Quark Matter: Precursors Of Chiral Transition At
  %Finite Temperature,''
  Phys.\ Rev.\ Lett.\  {\bf 55}, 158 (1985).
  %%CITATION = PRLTA,55,158;%%

%\cite{Shuryak:2003ty}
\bibitem{Shuryak:2003ty}
  E.~V.~Shuryak and I.~Zahed,
  %``Rethinking the properties of the quark gluon plasma at T approx. T(c),''
  Phys.\ Rev.\  C {\bf 70}, 021901 (2004);
%  [arXiv:hep-ph/0307267],
  %%CITATION = PHRVA,C70,021901;%%
%\bibitem{Shuryak:2004tx}
%  E.~V.~Shuryak and I.~Zahed,
  %``Towards a theory of binary bound states in the quark gluon plasma,''
  Phys.\ Rev.\  D {\bf 70}, 054507 (2004).
%  [arXiv:hep-ph/0403127].
  %%CITATION = PHRVA,D70,054507;%%

%\bibitem{GellMann:1964nj}
%  M.~Gell-Mann,
%  %``A Schematic Model Of Baryons And Mesons,''
%  Phys.\ Lett.\  {\bf 8}, 214 (1964);
%  %%CITATION = PHLTA,8,214;%%
%%\bibitem{Ida:1966ev}
%  M.~Ida and R.~Kobayashi,
%  %``Baryon resonances in a quark model,''
%  Prog.\ Theor.\ Phys.\  {\bf 36} (1966) 846;
%  %%CITATION = PTPKA,36,846;%%
%%\bibitem{Lichtenberg:1981pp}
%  D.~B.~Lichtenberg and L.~J.~Tassie,
%  %``Baryon Mass Splitting in a Boson-Fermion Model,"
%  Phys.\ Rev.\  {\bf 155} (1967) 1601;
%  D.~B.~Lichtenberg, L.~J.~Tassie, and P.~J.~Kelemen,
%%  %``Quark-Diquark Model of Baryons and SU(6),"
%  Phys.\ Rev.\  {\bf 167} (1968) 1535;
%  D.~B.~Lichtenberg,
%%  %``Baryon Supermultiplets Of SU(6) X O(3) In A Quark-Diquark Model,''
%  Phys.\ Rev.\  {\bf 178} (1969) 2197.
%%  %%CITATION = PHRVA,178,2197;%%

%\bibitem{Lichtenberg:1982jp}
%   D.~B.~Lichtenberg, W.~Namgung, E.~Predazzi and J.~G.~Wills,
%  %``Baryon Masses In A Relativistic Quark - Diquark Model,''
%  Phys.\ Rev.\ Lett.\  {\bf 48}, 1653 (1982).
%  %%CITATION = PRLTA,48,1653;%%

%\bibitem{Semay:1994ht}
%  C.~Semay and B.~Silvestre-Brac,
%  %``Diquonia and potential models,''
%  Z.\ Phys.\  C {\bf 61}, 271 (1994).
%  %%CITATION = ZEPYA,C61,271;%%

%\bibitem{Santopinto:2004hw}
%  E.~Santopinto,
%  %``An interacting quark-diquark model of baryons,''
%  Phys.\ Rev.\  C {\bf 72}, 022201 (2005).
%%  [arXiv:hep-ph/0412319].
%  %%CITATION = PHRVA,C72,022201;%%

%\bibitem{Jaffe:1976ig}
%  R.~L.~Jaffe,
%  %``Multi-Quark Hadrons. 1. The Phenomenology Of (2 Quark 2 Anti-Quark)
%  %Mesons,''
%  Phys.\ Rev.\  D {\bf 15}, 267 (1977);
%  %%CITATION = PHRVA,D15,267;%%
%%\bibitem{Jaffe:2003sg}
%  R.~L.~Jaffe and F.~Wilczek,
%  %``Diquarks and exotic spectroscopy,''
%  Phys.\ Rev.\ Lett.\  {\bf 91}, 232003 (2003).
%%  [arXiv:hep-ph/0307341].
%  %%CITATION = PRLTA,91,232003;%%

%\bibitem{Ebert:1995fp}
%   D.~Ebert, T.~Feldmann, C.~Kettner and H.~Reinhardt,
%  %``A Diquark model for baryons containing one heavy quark,''
%  Z.\ Phys.\  C {\bf 71}, 329 (1996).
%%  [arXiv:hep-ph/9506298].
%  %%CITATION = ZEPYA,C71,329;%%

%%\cite{Nagata:2008hi}
%\bibitem{Nagata:2008hi}
%  K.~Nagata and A.~Hosaka,
%  %``Structure of the Nucleon and Roper Resonance with Diquark Correlations,''
%  arXiv:0802.2366 [hep-ph].
%  %%CITATION = ARXIV:0802.2366;%%

%\bibitem{Alexandrou:2006cq}
%  C.~Alexandrou, Ph.~de Forcrand and B.~Lucini,
%  %``Evidence for diquarks in lattice QCD,''
%  Phys.\ Rev.\ Lett.\  {\bf 97}, 222002 (2006).
%%  [arXiv:hep-lat/0609004].
%  %%CITATION = PRLTA,97,222002;%%

%%\cite{Koch:2005vg}
%\bibitem{Koch:2005vg}
%   V.~Koch, A.~Majumder and J.~Randrup,
%  %``Baryon-strangeness correlations: A diagnostic of strongly interacting
%  %matter,''
%  Phys.\ Rev.\ Lett.\  {\bf 95}, 182301 (2005).
%%  [arXiv:nucl-th/0505052].
%  %%CITATION = PRLTA,95,182301;%%

%\cite{Alford:1997zt}
%\bibitem{Alford:1997zt}
%   M.~G.~Alford, K.~Rajagopal and F.~Wilczek,
%  %``QCD at finite baryon density: Nucleon droplets and color
%  %superconductivity,''
%  Phys.\ Lett.\  B {\bf 422}, 247 (1998);
%%  [arXiv:hep-ph/9711395].
%  %%CITATION = PHLTA,B422,247;%%
%%%\cite{Rapp:1997zu}
%%\bibitem{Rapp:1997zu}
%    R.~Rapp, T.~Schafer, E.~V.~Shuryak and M.~Velkovsky,
%%  %``Diquark Bose condensates in high density matter and instantons,''
%  Phys.\ Rev.\ Lett.\  {\bf 81}, 53 (1998).
%%  [arXiv:hep-ph/9711396].
%%  %%CITATION = PRLTA,81,53;%%

%\bibitem{Nishida:2005ds}
%  Y.~Nishida and H.~Abuki,
%  %``BCS-BEC crossover in relativistic superfluid and its possible  realization
%  %in QCD,''
%  Phys.\ Rev.\  D {\bf 72}, 096004 (2005).
%%  [arXiv:hep-ph/0504083].
%  %%CITATION = PHRVA,D72,096004;%%

%\cite{Lee:2007wr}
\bibitem{Lee:2007wr}
  S.~H.~Lee, K.~Ohnishi, S.~Yasui, I.~K.~Yoo and C.~M.~Ko,
  %``\Lambda_{c} enhancement from strongly coupled quark-gluon plasma,''
  arXiv:0709.3637 [nucl-th].
  %%CITATION = ARXIV:0709.3637;%%

%\bibitem{De Rujula:1975ge}
%   A.~De Rujula, H.~Georgi and S.~L.~Glashow,
%  %``Hadron Masses In A Gauge Theory,''
%  Phys.\ Rev.\  D {\bf 12}, 147 (1975).
%  %%CITATION = PHRVA,D12,147;%%

%\bibitem{Oka:1989ud}
%  M.~Oka and S.~Takeuchi,
%  %``INSTANTON INDUCED QUARK QUARK INTERACTIONS IN TWO BARYON SYSTEMS,''
%  Phys.\ Rev.\ Lett.\  {\bf 63}, 1780 (1989);
%  %%CITATION = PRLTA,63,1780;%%
%%\bibitem{Oka:1990vx}
%%  M.~Oka and S.~Takeuchi,
%  %``Instanton Induced Interaction And The Strange Dibaryons,''
%  Nucl.\ Phys.\  A {\bf 524}, 649 (1991).
%  %%CITATION = NUPHA,A524,649;%%

\bibitem{Jaffe:2004ph}
  R.~L.~Jaffe,
  %``Exotica,''
  Phys.\ Rept.\  {\bf 409}, 1 (2005);
%  [Nucl.\ Phys.\ Proc.\ Suppl.\  {\bf 142}, 343 (2005)];
%  [arXiv:hep-ph/0409065].
%  %%CITATION = NUPHZ,142,343;%%
%%\cite{Jaffe:2005md}
%%\bibitem{Jaffe:2005md}
% R.~Jaffe,
  %``Color non-singlet spectroscopy,''
 Phys.\ Rev.\  D {\bf 72}, 074508 (2005).
%%  [arXiv:hep-ph/0507149].
%  %%CITATION = PHRVA,D72,074508;%%

%\bibitem{Yasui:2007dv}
%  S.~Yasui and M.~Oka,
%  %``Triquark structure and isospin symmetry breaking in exotic Ds mesons,''
%  Phys.\ Rev.\  D {\bf 76}, 034009 (2007).
%%  [arXiv:0704.1345 [hep-ph].
%  %%CITATION = ARXIV:0704.1345;%%

\bibitem{Voloshin03}
  D.~Molnar and S.~A.~Voloshin,
  %``Elliptic flow at large transverse momenta from quark coalescence,''
 Phys.\ Rev.\ Lett.\  {\bf 91}, 092301 (2003).
%  [arXiv:nucl-th/0302014].
  %%CITATION = PRLTA,91,092301;%%

\bibitem{Hwa:2002tu}
  R.~C.~Hwa and C~.B.~Yang,
  %``Scaling behavior at high p(T) and the p/pi ratio,''
  Phys.\ Rev.\  C {\bf 67}, 034902 (2003).
%  [arXiv:nucl-th/0211010].
  %%CITATION = PHRVA,C67,034902;%%

\bibitem{greco}
%\cite{Greco:2003xt}
%\bibitem{Greco:2003xt}
  V.~Greco, C.~M.~Ko and P.~L\'evai,
  %``Parton coalescence and antiproton/pion anomaly at RHIC,''
  Phys.\ Rev.\ Lett.\  {\bf 90}, 202302 (2003);
%  [arXiv:nucl-th/0301093].
  %%CITATION = PRLTA,90,202302;%%
%\cite{Greco:2003mm}
%\bibitem{Greco:2003mm}
%  V.~Greco, C.~M.~Ko and P.~Levai,
  %``Parton coalescence at RHIC,''
  Phys.\ Rev.\  C {\bf 68}, 034904 (2003).
%  [arXiv:nucl-th/0305024].
  %%CITATION = PHRVA,C68,034904;%%
%\cite{Greco:2003vf}
%\bibitem{Greco:2003vf}
%\sout{  V.~Greco {\it et al.}, }
%  V.~Greco, C.~M.~Ko and R.~Rapp,
  %``Quark coalescence for charmed mesons in ultrarelativistic heavy-ion
  %collisions,''
%\sout{  Phys.\ Lett.\  B {\bf 595}, 202 (2004). }
%  [arXiv:nucl-th/0312100].
  %%CITATION = PHLTA,B595,202;%%

\bibitem{Fries:2003vb}
    R.~J.~Fries, B.~Muller, C.~Nonaka and S.~A.~Bass,
  %``Hadronization in heavy ion collisions: Recombination and fragmentation  of
  %partons,''
  Phys.\ Rev.\ Lett.\  {\bf 90}, 202303 (2003).
%  [arXiv:nucl-th/0301087].
  %%CITATION = PRLTA,90,202303;%%

\bibitem{greco1} V. Greco and C. M. Ko, Phys. Phys. Rev. C {\bf 70},
024901, (2004).

\bibitem{kolb}P. R. Kolb {\it et al.},
%L. W. Chen, V. Greco, and C. M. Ko,
Phys. Rev. C {\bf 69}, 051901(R) (2004).

\bibitem{greco2}V. Greco {\it et al.},
%C. M. Ko, and R. Rapp,
Phys. Lett. B {\bf 595}, 202 (2004).

\bibitem{chen}
   L.~W.~Chen, C.~-M.~Ko, W.~Liu, and M.~Nielsen,
  %``$D_{sJ}$(2317) meson production at RHIC,''
Phys.\ Rev.\ C {\bf 76}, 014906 (2007).
  %%CITATION = HEP-PH/0703071;%%

\bibitem{Chen:2003tn}
    L.~W.~Chen, V.~Greco, C.~M.~Ko, S.~H.~Lee and W.~Liu,
  %``Pentaquark baryon production at the Relativistic Heavy Ion Collider,''
  Phys.\ Lett.\  B {\bf 601}, 34 (2004).
%  [arXiv:nucl-th/0308006].
  %%CITATION = PHLTA,B601,34;%%

%\cite{Zhang:2007dm}
\bibitem{Zhang:2007dm}
   B.~W.~Zhang, C.~M.~Ko and W.~Liu,
  %``Thermal Charm Production in Quark-Gluon Plasma at LHC,''
  arXiv:0709.1684 [nucl-th].
  %%CITATION = ARXIV:0709.1684;%%

%%\cite{Adler:2005ab}
%\bibitem{Adler:2005ab}
%  S.~S.~Adler {\it et al.}  [PHENIX Collaboration],
%  %``Measurement of single electron event anisotropy in Au + Au collisions  at
%  %s(NN)**(1/2) = 200-GeV,''
%  Phys.\ Rev.\  C {\bf 72}, 024901 (2005).
%%  [arXiv:nucl-ex/0502009].
%  %%CITATION = PHRVA,C72,024901;%%

%%\bibitem{laue} F.~Laue [STAR Collaboration], J.\ Phys.\ G {\bf 31},
%%S27 (2005).

%%\cite{Laue:2004tf}
%\bibitem{Laue:2004tf}
%  F.~Laue  [STAR Collaboration],
%  %``Single electron elliptic flow measurements in Au + Au collisions from
%  %STAR,''
%  J.\ Phys.\ G {\bf 31}, S27 (2005).
%%  [arXiv:nucl-ex/0411007].
%  %%CITATION = JPHGB,G31,S27;%%

%\cite{van Hees:2004gq}
\bibitem{van Hees:2004gq}
  H.~van Hees and R.~Rapp,
  %``Thermalization of heavy quarks in the quark-gluon plasma,''
  Phys.\ Rev.\ C {\bf 71}, 034907 (2005).
%  [arXiv:nucl-th/0412015].
  %%CITATION = PHRVA,C71,034907;%%

%\cite{Zhang:2005ni}
\bibitem{Zhang:2005ni}
  B.~Zhang, L.~W.~Chen and C.~M.~Ko,
  %``Charm elliptic flow at RHIC,''
  Phys.\ Rev.\  C {\bf 72}, 024906 (2005);
%  [arXiv:nucl-th/0502056].
  %%CITATION = PHRVA,C72,024906;%%
%\cite{Zhang:2005nt}
%\bibitem{Zhang:2005nt}
%  B.~Zhang, L.~W.~Chen and C.~M.~Ko,
  %``Charm elliptic flow in Au + Au collisions at RHIC,''
  Nucl.\ Phys.\  A {\bf 774}n charm production in nuclear collisions at SPS/FAIR
  %energies and the possible influence of a hot hadronic medium,''
  arXiv:0708.1488 [nucl-th].
  %%CITATION = ARXIV:0708.1488;%%

%\cite{Molnar:2004ph}
\bibitem{Molnar:2004ph}
  D.~Molnar,
  %``Charm elliptic flow from quark coalescence dynamics,''
  J.\ Phys.\ G {\bf 31}, S421 (2005).
%  [arXiv:nucl-th/0410041].
  %%CITATION = JPHGB,G31,S421;%%

\bibitem{Lee:2007tn}
   S.~H.~Lee, S.~Yasui, W.~Liu and C.~M.~Ko,
  %``Charmed exotics in Heavy Ion Collisions,''
  arXiv:0707.1747 [hep-ph].
  %%CITATION = ARXIV:0707.1747;%%

\bibitem{Kushnirenko:2000ed}
  A.~Kushnirenko {\it et al.}  [SELEX Collaboration],
  %``Precision measurements of the Lambda/c+ and D0 lifetimes,''
  Phys.\ Rev.\ Lett.\  {\bf 86}, 5243 (2001).
%  [arXiv:hep-ex/0010014].
  %%CITATION = PRLTA,86,5243;%%

\bibitem{Yao:2006px}
  W.~M.~Yao {\it et al.}  [Particle Data Group],
  %``Review of particle physics,''
  J.\ Phys.\ G {\bf 33}, 1 (2006).
  %%CITATION = JPHGB,G33,1;%%

\bibitem{PBM07}
  A.~Andronic {\it et al.},
  %``Charmonium and open charm production in nuclear collisions at SPS/FAIR
  %energies and the possible influence of a hot hadronic medium,''
  arXiv:0708.1488 [nucl-th].
  %%CITATION = ARXIV:0708.1488;%%

%%\cite{Liu:2001ce}
%\bibitem{Liu:2001ce}
%   W.~Liu, C.~M.~Ko and Z.~W.~Lin,
%  %``Cross section for charmonium absorption by nucleons,''
%  Phys.\ Rev.\  C {\bf 65}, 015203 (2002);
%  %%CITATION = PHRVA,C65,015203;%%
%%\cite{Oh:2007ej}
%%\bibitem{Oh:2007ej}
%  Y.~Oh {\it et al.},
%%  Y.~Oh, W.~Liu and C.~M.~Ko,
%  %``J/psi absorption by nucleons in the meson-exchange model,''
%%  Phys.\ Rev.\  C {\bf 75}, 064903 (2007).
%%  [arXiv:nucl-th/0702077].
%  %%CITATION = PHRVA,C75,064903;%%

%\bibitem{Matsui:1986dk}
%  T.~Matsui and H.~Satz,
  %``J/psi Suppression by Quark-Gluon Plasma Formation,''
%  Phys.\ Lett.\  B {\bf 178}, 416 (1986).
  %%CITATION = PHLTA,B178,416;%%

% \bibitem{ALICE} http://aliceinfo.cern.ch/
\bibitem{alice1}
   ``ALICE Technical Design Report of the Inner Tracking System (ITS)",
   CERN /LHCC 99-12, ALICE TDR 4, 1999

%\bibitem{STAR} http://www.star.bnl.gov/
\bibitem{STAR1} http://www.star.bnl.gov/STAR/central/presentations
    /2007/rhicags/Thomas\_Jim.pdf

\bibitem{PHENIX1} http://www.phenix.bnl.gov/phenix/WWW/publish/adion
    /HEP2007/Alan\_Dion\_HEP2007.pdf

\bibitem{vogt1} M. Cacciari, P. Nason, and R. Vogt, Phys. Rev. Lett.
{\bf 95}, 122001 (2005).

\bibitem{vogt2} A. Accardi {\it et al.}, hep-ph/0308248, p.79.

%\bibitem{adler} S.~S.~Adler {\it et al.} (PHENIX Collaboration),
%Phys.\ Rev.\ C {\bf 72}, 024901 (2005).

%\bibitem{zhang} B. Zhang, L. W. Chen, and C. M. Ko, Phys. Rev. C
%{\bf 72}, 024906 (2005); Nucl. Phys. A {\bf 774}, 665 (2006).

%\bibitem{hees} H. van Hees and R. Rapp, Phys. Rev. C {\bf 71},
%034907 (2005).

%\bibitem{molnar} D. Molnar, J. Phys. G {\bf 31}, S421 (2005).

%\bibitem{KanadaEn'yo:2006zk}
%  Y.~Kanada-En'yo and B.~Muller,
  %``Suppression of p-wave baryons in quark recombination,''
%  Phys.\ Rev.\  C {\bf 74}, 061901 (2006).
%  [arXiv:nucl-th/0608015].
  %%CITATION = PHRVA,C74,061901;%%

\end{thebibliography}
\end{document}